\documentclass[final]{cvpr}

\usepackage{times}
\usepackage{epsfig}
\usepackage{graphicx}
\usepackage{amsmath}
\usepackage{amssymb}
\usepackage{subfig}
\usepackage{multirow}
\usepackage{tabularx,booktabs}

\usepackage[pagebackref=true,breaklinks=true,colorlinks,bookmarks=false]{hyperref}
\def \img {\mathbf{I}}

\newcolumntype{Y}{>{\centering\arraybackslash}X}

\def\cvprPaperID{5944} 
\def\confYear{CVPR 2021}

\begin{document}

\title{Exploiting Aliasing for Manga Restoration}

\author{Minshan Xie$^{1,2,}$\footnotemark[1]~ \quad Menghan Xia$^{1,}$\footnotemark[1]~ \quad Tien-Tsin Wong$^{1,2,}$\footnotemark[2]~\\
$^1$~Department of Computer Science and Engineering, The Chinese University of Hong Kong\\
$^2$~Shenzhen Key Laboratory of Virtual Reality and Human Interaction Technology, SIAT, CAS\\
{\tt\small \{msxie, mhxia, ttwong\}@cse.cuhk.edu.hk}}
\maketitle
\renewcommand{\thefootnote}{\fnsymbol{footnote}}
\footnotetext[1]{Equal contributions.}
\footnotetext[2]{Corresponding author.}

\pagestyle{empty}
\thispagestyle{empty}

\begin{abstract}
As a popular entertainment art form, manga enriches the line drawings details with bitonal screentones. However, manga resources over the Internet usually show screentone artifacts because of inappropriate scanning/rescaling resolution. In this paper, we propose an innovative two-stage method to restore quality bitonal manga from degraded ones. Our key observation is that the aliasing induced by downsampling bitonal screentones can be utilized as informative clues to infer the original resolution and screentones. First, we predict the target resolution from the degraded manga via the Scale Estimation Network (SE-Net) with spatial voting scheme. Then, at the target resolution, we restore the region-wise bitonal screentones via the Manga Restoration Network (MR-Net) discriminatively, depending on the degradation degree. Specifically, the original screentones are directly restored in pattern-identifiable regions, and visually plausible screentones are synthesized in pattern-agnostic regions. Quantitative evaluation on synthetic data and visual assessment on real-world cases illustrate the effectiveness of our method.
\end{abstract}

\section{Introduction}
\label{sec:intro}

\begin{figure}
    \centering
    \includegraphics[width=0.95\linewidth]{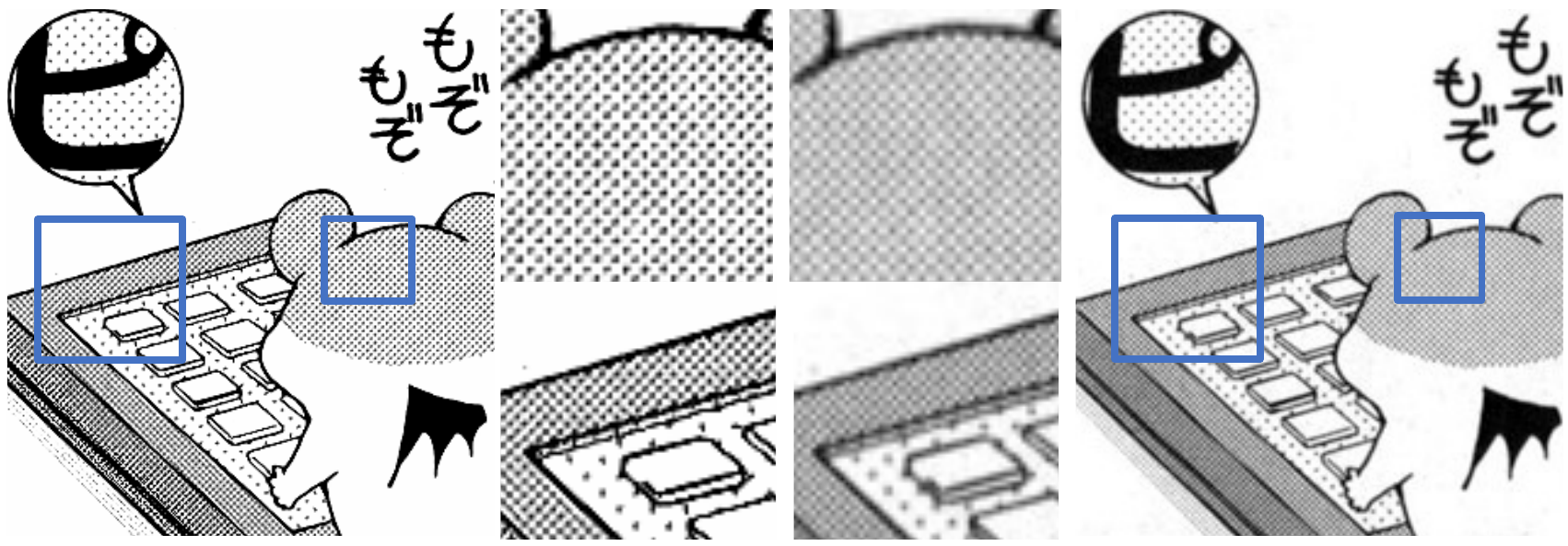}\\
    (a) Standard manga \hspace{.2cm} (b) Blow-up \hspace{.2cm} (c) Degraded manga\\
    \vspace{-.5em}
    \caption{The screentones in the manga image with insufficient resolution are blurry while the desired screentones should be sharply bitonal. The image comes from the Manga109 dataset \cite{mtap_matsui_2017}. Akuhamu \textcopyright Arai Satoshi}\vspace{-0.5em}
    \label{fig:intro}
\end{figure}

Manga, also known as Japanese comics, is a popular entertainment art form. 
One of the key differences between manga and other illustration types is the use of \textit{screentones}, regular or stochastic black-and-white patterns to render intensity, textures and shadings (Figure~\ref{fig:intro}(a)). Although furnishing impressive visual impact, the existence of screentones makes it tricky to resample manga images.
For instance, when being undersampled, the bitonal regular patterns may get ruined and present incoherent visual effects (Figure~\ref{fig:intro}(c)). Unfortunately, it is common to see such screentone artifacts from the manga images over the Internet (e.g. Manga109~\cite{mtap_matsui_2017}), probably due to the poor scanners or storage limitation in the old days. In this background, we are motivated to restore these low-quality legacy mangas and show their original appearances.

\begin{figure}
    \centering
    \includegraphics[width=.97\linewidth]{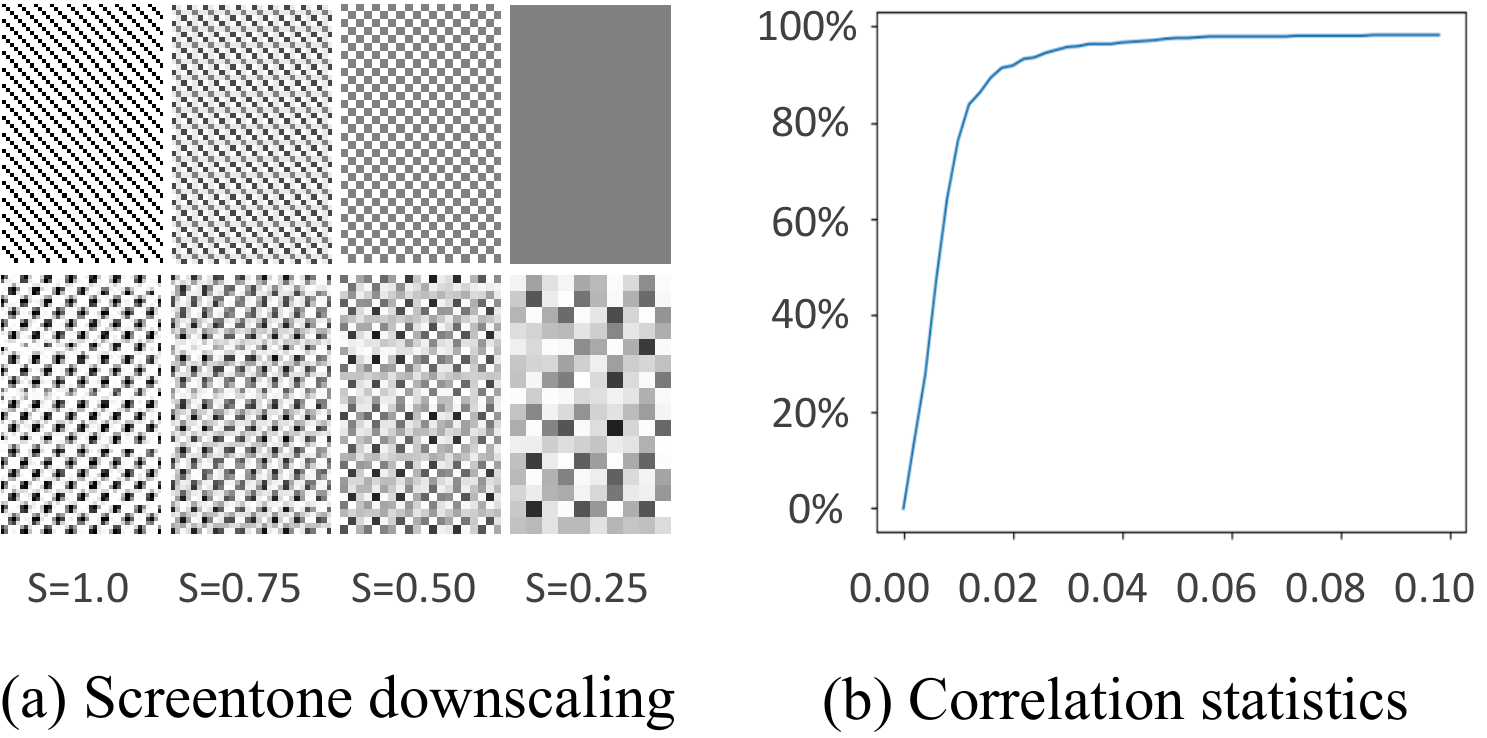}
    \\ \vspace{-.5em}
    \caption{Our observation. (a) Aliasing from screentone downscaling is visually distinctive, depending on the scale factor and screentone type. (b) Statistic of scale prediction error on a synthetic dataset (degraded scale ranges $1.0\sim4.0$). A prediction error below $0.02$ is achieved over 91\% samples, indicating the strong correlation between aliasing property and the applied downscaling factors.} 
    \label{fig:observation}
\end{figure}

Unlike natural images dominating with low-frequency components, manga images mainly consist of regular high-frequency patterns that are pickier at the representing resolution. Specifically, for a quality bitonal manga image of resolution $T$, it is generally impossible to present the screentones in a both bitonal and perceptually consistent manner on the resolution $S \neq T_k \in \{kT | k=1,2,3,...,n\}$. That means, to restore a manga image, we first need to figure out a target resolution that is able to present the potential target screentones, and then restore the screentones from the degraded ones at that scale. Apparently, this tricky requirement excludes the feasibility of existing Single Image Super-Resolution (SISR) methods~\cite{dong2015image,ge2018image,hu2019meta} and image restoration methods \cite{liu2018non,zhang2017learning,dong2015compression,tao2018scale}.
Instead, our key idea is inspired by an interesting observation that the aliasing caused by downsampling the bitonal screentones is usually distinctive on the downscaling factor and screentone type, as illustrated in Figure~\ref{fig:observation}. These may serve as informative clues to infer the original screentones and their associated resolution.


To this end, we propose an innovative two-stage manga restoration method. First, we utilize the Scale Estimation Network (SE-Net) to predict the target resolution from the degraded screentones. There are usually multiple screentones within a single manga image, and each of them may contribute differently to the prediction accuracy. This is effectively tackled through our proposed spatial voting scheme based on confidence.
At the predicted resolution scale, we restore the region-wise bitonal screentones via the Manga Restoration Network (MR-Net). Considering the different degradation degrees, the manga image is restored with two parallel branches: the original screentones are restored for pattern-identifiable regions, and random screentones are synthesized under intensity conformity for pattern-agnostic regions. Specifically, this region-wise classification is determined adaptively through a learned confidence map.
Separately, the two networks are trained over the mixture dataset of synthetic manga and real ones, in a semi-supervised manner. 

We have evaluated our method on our synthetic testset and some real-world cases. Quantitative evaluation demonstrates that our restored manga images achieve high PSNR and SSIM on synthetic data. Meanwhile, qualitative evaluation of real-world cases evidences the potential for practical usage. 
In summary, this paper has the contributions:

\begin{itemize}
    \item The first manga restoration method that restores the bitonal screentones at a learned resolution.
    \item A manga restoration network that restores the region-wise screentones adaptively based on a learned confidence map.
\end{itemize}
While our current method is tailored for the manga restoration problem, our proposed framework has the potential to be extended to natural images containing the regular textures. For example, the checkerboard artifact resulted from undersampling the regular textures should share a similar property as the screentones. 

\section{Related Work}
\label{sec:related_work}
\subsection{Manga Screening}
Attempts have been made to generate screentone manga automatically from grayscale/color images or line drawings. 
Qu et al. \cite{qu-2008-richness} applied a variety of screentones to segments based on the similarity between texture, color, and tone to preserve the visual richness. However, the method failed to restore bitonal manga from the degraded version as the degraded screentones maybe significantly differ from the original patches. 
Li et al. \cite{li2017deep} presented an effective way to \emph{synthesize} screen-rich manga from line drawings. 
Tsubota et al.\cite{tsubota2019synthesis} synthesize manga images by generating pixel-wise screentone class labels and further laying the corresponding screentones from database. However, these methods are highly dependent on the screentone set and cannot generate the original bitonal screentones. 
In contrast, our method attempts to recover the original version of the degraded manga by learning the degradation rules of screentones with generalization to real-world cases.

\begin{figure*}[ht]
    \centering
    \includegraphics[width=.95\linewidth]{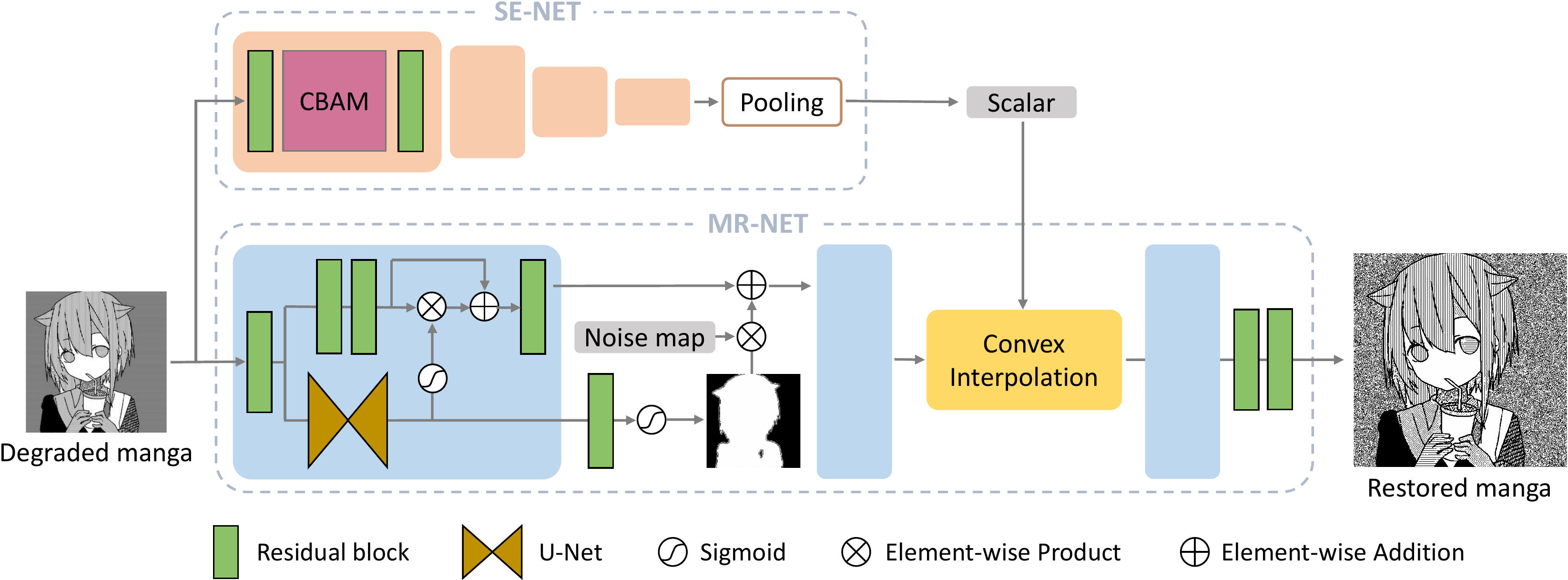}\vspace{-0.5em}
    \caption{System overview. Given the degraded manga image, the SE-Net first estimates the scalar that is required to upscale, and then the MR-Net restores the manga image at the predicted resolution scale.}
    \label{fig:overview}
\end{figure*}

\subsection{Single Image Super-Resolution}
As a classic vision task, Single Image Super-Resolution (SISR) aims at reconstructing the high-resolution (HR) version from the low-resolution (LR) images. Traditional methods mainly leverage dictionary learning~\cite{yang2008image} or database retrieval\cite{chang2004super,freeman2002example,timofte2013anchored} to reconstruct the high-frequency details for the low-resolution input. However, due to the limited representation capability of hand-crafted features and lack of semantic level interpretation, these methods struggle to achieve photorealistic results.

Recently, as deep learning techniques on the rise, the state-of-the-art SISR has been updated continuously by these data-driven approaches.
Given pairs of LR and HR images, some studies ~\cite{dong2015image} attempt to solve it as a regression problem that maps LR images to their corresponding HR images. Many follow-ups reached a more accurate HR image by designing better network architectures, such as VDSR \cite{kim2016accurate}, SRResNet\cite{ledig2017photo}, LapSRN \cite{lai2018fast}, or more powerful loss functions, like EDSR\cite{lim2017enhanced}. 
However, these methods tend to generate blurry results as they failed to recover the lost high-frequency signal that has little correlation with the LR image.  
To recover these lost details, some approaches ~\cite{ledig2017photo,ge2018image,wang2018esrgan} adopt generative adversarial networks (GANs)~\cite{goodfellow2014generative} to generate stochastic details.
SRGAN~\cite{ledig2017photo} attempts to recover the lost details by adopting a discriminator to tell what kind of high-frequency details look natural. 
As a step further, a super-resolution method of arbitrary scale~\cite{hu2019meta} is proposed to reconstruct the HR image with continuous scale factor, which is the most related work to our method.
However, all these methods, mainly working on natural images, never consider the scale suitability when recovering the high-resolution images. This is just the intrinsic difference between SISR and our problem. Indeed, our method attempts to first obtain a proper resolution from the degraded screentones which then helps to restore the bitonal nature. 
\section{Scale-Awared Manga Restoration}
\label{sec:approach}

Given a degraded manga image, we aim to restore the bitonal screentones to be as conformable as possible to the original version. As illustrated in Figure~\ref{fig:overview}, it is formulated by a two-stage restoration framework including restorative scale estimation and manga screentone restoration. The detailed network architectures are provided in the supplementary material.

\subsection{Problem Formulation}
\label{subsec:problem_formula}

Let $\img_{\rm gt}$ be the original bitonal manga image. Generally, the degraded image $\img_x$ can be modeled as the output of the following degradation:
\begin{equation}
\img_x = (\img_{\rm gt} \otimes\kappa)\downarrow s_{\rm gt} + \mathbf{N}_{\varsigma},
\end{equation}
where $\{\kappa, s_{\rm gt}, \varsigma\}$ parameterizes the degradation process. $\img_{\rm gt}\otimes\kappa$ denotes the convolution between a blur kernel $\kappa$ and the image $\img_{\rm gt}$, $\downarrow s_{\rm gt}$ denotes the downsampling operation with the scale factor $s_{\rm gt}$. Without losing generality, $\mathbf{N}_{\varsigma}$ describes other potential noises induced by scanning process or lossy image format like JPEG, which overall is modelled as additive noises with standard deviation $\varsigma$.

This seems very similar to the formula adopted in super-resolution~\cite{wang2020deep}, but the problem is crucially different due to the special nature of manga restoration. As introduced in Section~\ref{sec:intro}, it is impossible to recover the original bitonal screentones from degraded ones unless it is represented with appropriate resolution. To tackle this problem, we need to: (i) figure out the target resolution by estimating the desired scale factor: $s_y=g(\img_x) \rightarrow s_{\rm gt}$; (ii) perform the manga screentone restoration at the estimated resolution, which is a comprehensive deconvolution and denoising process: $\img_y=f(\img_x, s_y) \rightarrow \img_{\rm gt}$. It makes sense because of the distinctive correlation between $\img_{\rm gt}$ and $\img_x$ that conditions on $s_{\rm gt}$ and the screentone type, as observed in Figure~\ref{fig:observation}. In particular, we try to model the functions $g(\cdot)$ and $f(\cdot)$ by training two neural networks respectively.

\subsection{Restorative Scale Estimation}
\label{subsec:scale_estimation}

Based on the degraded manga image, we utilize the Scale Estimation Network (SE-Net) to estimate the downscaling scalar that has been applied to the original bitonal manga image. This is a prerequisite of the subsequent manga restoration that requires a screentone-dependent restorative resolution.

\begin{figure}[!t]
    \centering
    \subfloat[Bitonal manga]{\includegraphics[width=.27\linewidth]{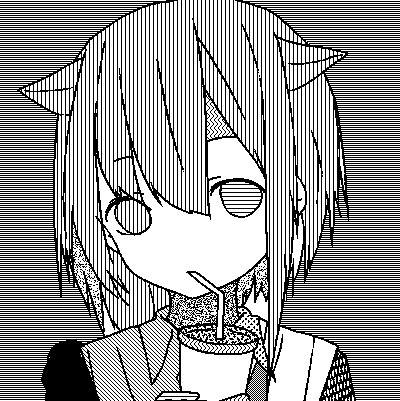}}\hspace{.02in}
    \subfloat[98\%]{\includegraphics[width=.27\linewidth]{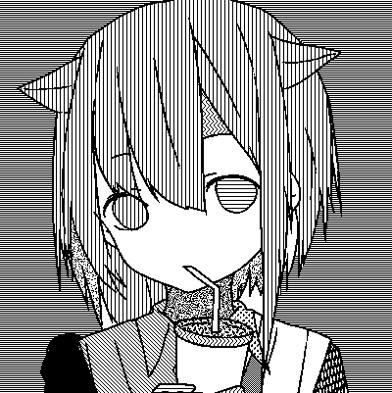}}\hspace{.02in}
    \subfloat[83\%]{\includegraphics[width=.27\linewidth]{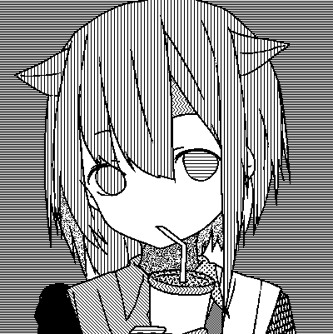}}\\ \vspace{-0.1in}
    \subfloat[67\%]{\includegraphics[width=.27\linewidth]{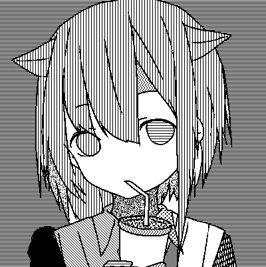}}\hspace{.02in}
    \subfloat[50\%]{\includegraphics[width=.27\linewidth]{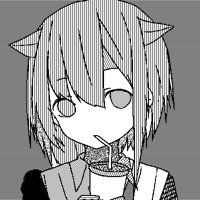}}\hspace{.02in}
    \subfloat[25\%]{\includegraphics[width=.27\linewidth]{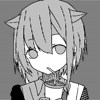}}\\ \vspace{-0.5em}
    \caption{Manga degradation with different downscaling factors. Different screentones will have different degradations with the same downscaling factor. The screentones on background are degraded to plain region which gives no clue for restoration with 50\% resolution while the screentones on foreground still retains informative patterns.}\vspace{-0.5em}
    \label{fig:degrade}
\end{figure}

\paragraph{Scale Estimation Network (SE-Net).}
The SE-Net takes the degraded image $\img_x$ as input and outputs the estimated scale factor $s_y$ for further restoration.
Figure~\ref{fig:overview} shows the abstract structure of the SE-Net, which cascades four downsample modules and an adaptive pooling layer. As a common case, a single manga image contains multiple screentone regions and each of them degrades to some different extent depending on the screentone pattern types, as shown in Figure~\ref{fig:degrade}. Consequently, these screentone regions might be informative differently to infer the downscaling factor $s_{\rm gt}$, which motivates us to adopt the Convolutional Block Attention Module (CBAM)\cite{woo2018cbam} to focus on deterministic regions and ignore noisy ambiguous regions. 
Since the attention of CBAM is performed in the feature domain along both channel and spatial dimensions, the intermediate feature maps are adaptively optimized with sufficient flexibility along with these downsample modules. 
Besides, to get the scalar output, we perform the adaptive pooling on the feature maps, where a global spatial pooling and a fully-connected layer are combined.  

\paragraph{Loss Function.}
The SE-Net is trained with the loss function comprised of two terms: scale loss $\mathcal{L}_{\rm scl}$ and consistency loss $\mathcal{L}_{\rm cons}$.
\begin{equation}
\mathcal{L}_{\rm SE} = \mathcal{L}_{\rm scl} + \alpha \mathcal{L}_{\rm cons},
\end{equation}
where $\alpha=0.1$ balances the magnitude of the two terms.

\textit{Scale loss.}
Given degraded image $\img_x$, the scale loss $\mathcal{L}_{\rm scl}$ is to encourage the SE-Net to generate a scale factor $s_y$ which is as close as possible to the ground truth $s_{\rm gt}$. 
\begin{equation}
\mathcal{L}_{\rm scl} = \|s_y-s_{\rm gt}\|_1,
\end{equation}

\textit{Consistency loss.}
When only trained on synthetic data based on $\mathcal{L}_{\rm scl}$, we find that it cannot generalize well to real-world cases. For example, on a scanned manga book, the predicted scale factors for different images from the same volume or even for different patches from the same image can be substantially different. Thus, we further introduce a consistency loss $\mathcal{L}_{\rm cons}$ to enforce the SE-Net to generate a consistent scale factor for the patches from the same manga image. Actually, this loss term benefits in two aspects: on the one hand, it stabilizes the network training by further introducing extra supervision; on the other, it enables semi-supervised training on the mixture data of synthetic manga and real-world manga and thus promotes the generalization performance on real-world cases.
\begin{equation}
\mathcal{L}_{\rm cons} = \|s_y -\frac{1}{M}\sum_{i=1}^M s^i_y\|_1,
\end{equation}
where $s_y^i$ denotes the predicted scale factor from the $i$-th of $M$ patches cropped from the same degraded image $\img_x$.

\subsection{Discriminative Restoration}
\label{subsec:screentone_restoration}

Based on the estimated scale factor, we utilize the Manga Restoration Network (MR-Net) to restore the screentones for the target manga image. According to the screentone degradation degrees, the MR-Net restores the manga image discriminatively on different regions: reconstruct the original screentones for pattern-identifiable regions while synthesizing plausible screentones for pattern-agnostic regions.

\paragraph{Manga Restoration Network (MR-Net).}
The MR-Net takes the degraded manga image $\img_x$ and the desired scale factor $s_{\rm gt}$ as input, while output the confidence map $\mathbf{M}_c$ and the restored manga image $\img_y$.
Figure~\ref{fig:overview} shows the abstract structure of the MR-Net, which employs the Residual Attention Module (RAM)~\cite{wang2017residual} as backbone unit to capture the screentone clue and restore the screentone regularity. Specifically, the attention features of the first RAM are further transformed to a single-channel confidence map $\mathbf{M}_c$ that is used to selectively introduce noises to the feature maps. The intuition is that the output manga image will be generated through two paths implicitly, i.e. reconstruction path and synthesis path, and the random noises are injected to add external variation for the regions under the charge of the synthesis path.
The second RAM further prepare the features for spatial upsampling, which is implemented by the convex interpolation block~\cite{teed2020raft} with learned neighborhood interpolative coefficients. Specifically, we interpolate a target pixel from $N$ known neighboring pixels $\{p_1, p_2, ...,p_N\}$ by computing the weighted sum: $\sum_{i=1}^{N}\alpha_ip_i$, where $\sum_{i=1}^{N}\alpha_i=1$ and $\forall \alpha_i\geq 0$. 
Then, the upsampled feature maps are transformed to the restored manga image by the rest layers. 

\paragraph{Loss Function.}
The optimization objective of the MR-Net comprises five terms: pixel loss $\mathcal{L}_{\rm pix}$, confidence loss $\mathcal{L}_{\rm conf}$, binarization loss $\mathcal{L}_{\rm bin}$, intensity loss $\mathcal{L}_{\rm itn}$ and homogeneity loss $\mathcal{L}_{\rm hom}$, written as:
\begin{equation}
\mathcal{L}_{\rm MR}= \mathcal{L}_{\rm pix}+\phi\mathcal{L}_{\rm conf}+\omega\mathcal{L}_{\rm bin}+\kappa \mathcal{L}_{\rm itn} + \gamma\mathcal{L}_{\rm hom},
\end{equation}
The empirical coefficients $\phi=0.5$, $\omega=0.5$, $\kappa=0.5$ and $\gamma=0.02$ are used in our experiment.

\textit{Pixel loss.}
The pixel loss $\mathcal{L}_{\rm pix}$ ensures the restored manga image $\img_y$ to be as similar as possible with the ground truth $\img_{\rm gt}$ on those pattern-identifiable regions and helps the network to reconstruct the original bitonal image. Here, we measure their similarity with the Mean Absolute Error (MAE), as defined in
\begin{equation}
\mathcal{L}_{\rm pix}=\|\mathbf{M}_c\odot|\img_y-\img_{\rm gt}|\|_1,
\end{equation}
where $\odot$ denotes element-wise multiplication and $|\cdot|$ denotes the operation to take the element-wise absolute value. The loss attention mechanism avoids overfitting to low-confidence regions, potentially focusing less on ambiguous regions.

\textit{Confidence loss.} 
The confidence loss $\mathcal{L}_{\rm conf}$ encourages the model to extract as many pattern-identifiable regions as possible. Sometimes, it is quite ambiguous to visually detect whether certain screentone degradation is pattern-identifiable or not. Instead, we formulate it as a confidence map $\mathbf{M}_c$ that is learned adaptively. Based on the prior that most degraded screentones are restorable, we encourage the model to restore as much as possible screentones through
\begin{equation}
\mathcal{L}_{\rm conf} = 1.0-\|\mathbf{M}_c\|_1.
\end{equation}
Here, the confidence map $\mathbf{M}_c$ has 1 represent pattern-identifiable regions and 0 indicates pattern-agnostic regions.

\textit{Binarization loss.}
To generate manga with bitonal screentones, we introduce the binarization loss $\mathcal{L}_{\rm bin}$ to encourage the network to generate black-and-white pixels, which is defined as
\begin{equation}
\mathcal{L}_{\rm bin} = \|||\img_y-0.5|-0.5|\|_1.
\end{equation}

\textit{Intensity loss.} 
The intensity loss $\mathcal{L}_{\rm itn}$ ensures that the generated manga image $\img_y$ visually conforms to the intensity of the target image $\img_{\rm gt}$. According to the low-frequency pass filter nature of Human Visual System (HVS), we compute this loss as:
\begin{equation}
\mathcal{L}_{\rm itn} = \|{\rm G}(\img_y)-{\rm G}(\img_{\rm gt})\|_1, 
\end{equation}
where ${\rm G}$ is a Gaussian blur operator with the kernel size of $11\times 11$. 
Specially, when calculating the intensity loss on the real-world cases, we resize the input to the target resolution and further smooth it with a Gaussian filter, which is still qualified guidance to constrain the intensity similarity.
In practice, this loss benefits in two folds.
For the ambiguous regions that are insufficient to restore the original screentones, we can still leverage this loss term to generate screentones with similar tonal intensity. In addition, it allows the training on real-world manga data to promote generalization performance. 

\textit{Homogeneity loss.}
The screentone homogeneity loss aims to impose the homogeneity within each screentone region.  
With the previous loss, we observe that the restored manga images sometimes have inconsistent screentones even in the same homogeneous regions. To alleviate this problem, we encourage the screentone features within each region to be similar through the homogeneity loss $\mathcal{L}_{\rm hom}$.
Here, we measure the screentone difference in the domain of ScreenVAE map~\cite{xie2020manga} that represents the screentone pattern as a smooth and interpolatable 4D vector and enables the pixelwise metrics (e.g. MSE) to be effective. In particular, we formulate the homogeneity loss $\mathcal{L}_{\rm hom}$ as:
\begin{equation}
\mathcal{L}_{\rm hom} = \frac{1}{N} \sum_{i=1}^{N}\|{\rm SP}_i(\Phi(\img_y))-\mu({\rm SP}_i(\Phi(\img_y)))\|_2,
\end{equation}
where $\Phi(\cdot)$ extract the ScreenVAE map of a manga image, ${\rm SP}_i$ denotes the the $i$-th superpixel that is achieved by segmenting on the ground truth manga $\img_{\rm gt}$, and $\mu(\cdot)$ computes the mean value. 
\section{Experimental Results}
\label{sec:experiment}

\subsection{Implementation details}
\label{subsec:implementation}

\paragraph{Data Preparation.}
Manga109 \cite{mtap_matsui_2017} is a public manga dataset, containing a total of about 20000 pieces of degraded image. However, the resolution of the manga images is low. 
Currently, there is no high-resolution public manga dataset which we can directly use as ground truth.
Fortunately, Li et al.\cite{li2017deep} proposed an effective manga synthesis method, which fills in line drawings with diverse screentones. 
To prepare paired training data, we synthesized 3000 pieces of bitonal manga images with the resolution of $2048\times 1536$, and simulate various degradation through the random combination of: (i) downsampling with multiple scale factors $s\in[1.0,4.0]$; (ii) JPEG compression with different quality factors $q\in[50,100]$; (iii) Gaussian noise with varied standard deviation $\mathcal{N}(0.0,5.0\sim15.0)$. 

\paragraph{Training Scheme.}
To favor the model generalization on real-world cases, we apply a semi-supervised strategy to train the SE-Net and the MR-Net separately, i.e. both paired synthetic data and unpaired real-world data are used for training.
In particular, for the synthetic data, all the losses are used, i.e. $\mathcal{L}_{\rm scl}$ and $\mathcal{L}_{\rm cons}$ in the first stage, and $\mathcal{L}_{\rm pix}$, $\mathcal{L}_{\rm conf}$, $\mathcal{L}_{\rm bin}$, $\mathcal{L}_{\rm itn}$, and $\mathcal{L}_{\rm hom}$ in the second stage. 
For the Manga109 which has no ground truth available, only $\mathcal{L}_{\rm cons}$ (stage 1) and $\mathcal{L}_{\rm conf}$, $\mathcal{L}_{\rm bin}$, $\mathcal{L}_{\rm itn}$ (stage 2) are used. 

We trained the model using PyTorch framework \cite{paszke2017automatic} and trained on Nvidia TITANX GPUs. The network weights are randomly initialized using the method of \cite{he2015delving}. During training, the models are optimized by Adam solver \cite{kingma2014adam} with $\beta_1=0.9$ and $\beta_2=0.999$. The learning rate is initialized to 0.0001.

\subsection{Scale Estimation Accuracy}
\label{subsec:scale_accuracy}

We evaluate the accuracy of our Scale Estimation Network (SE-Net) on synthetic data with different rescaling ranges. As tabulated in Table~\ref{tab:acc}, the accuracy decreases as the scale factor increases, because lower-resolution generally means severe degradation and hence involves more ambiguity for information inference. The average accuracy of the whole range $[1,4]$ is 0.9896. In other words, for a degraded image with downscaling factor of $T$, our estimated scale factor is expected to be $(1\pm0.0104)T$. In addition, we study the effectiveness of the CBAM and our consistency loss respectively.
We construct baseline module of the CBAM by removing the attention block, resulting in a simple residual block. We can observe that the CBAM improves the performance obviously, since the attention mechanism facilitates the network to focus on the informative regions while ignoring ambiguous regions.
\begin{table}[!t]
    \centering
    \renewcommand{\tabcolsep}{5pt}
    \caption{Accuracy evaluation of the estimated scale factor on synthetic data.}
    \vspace{-1.0em}
    \begin{tabular}{c|c|c}
        \hline
        Methods & Upscaling Range & Estimation Accuracy \\
        \hline
        \multirow{4}{*}{w/o CBAM} 
        & \multirow{1}{*}{[1,2]} & 0.9550\\ \cline{2-3}
        & \multirow{1}{*}{(2,3]} & 0.9538\\ \cline{2-3}
        & \multirow{1}{*}{(3,4]} & 0.9514\\ \cline{2-3}
        & \multirow{1}{*}{[1,4]} & 0.9542 \\ 
        \hline
        \multirow{4}{*}{w CBAM} 
        & \multirow{1}{*}{[1,2]} & 0.9823 \\ \cline{2-3}
        & \multirow{1}{*}{(2,3]} & 0.9936 \\ \cline{2-3}
        & \multirow{1}{*}{(3,4]} & 0.9929 \\ \cline{2-3}
        & \multirow{1}{*}{[1,4]} & 0.9896 \\ 
        \hline
    \end{tabular}
    \label{tab:acc}
\end{table}

\begin{table}[!t]
    \centering
    \renewcommand{\tabcolsep}{2pt}
    \caption{Scale prediction on the images from the same volume that shares the same ground-truth scale factor.}
    \vspace{-1.0em}
    \begin{tabularx}{.49\textwidth}{c*{4}{|Y}}
        \hline
        & \multicolumn{2}{c}{Synthetic data ($s_{\rm gt}=2$)} & \multicolumn{2}{|c}{Real data}\\
        \cline{2-5}
        & $\mu(s_{\rm y})$ & $\sigma(s_{\rm y})$ & $\mu(s_{\rm y})$ & $\sigma(s_{\rm y})$ \\
        \hline
        w/o $\mathcal{L}_{\rm cons}$ & 2.0483 & 0.1893 & 1.7995 & 0.4247 \\
        \hline
        w $\mathcal{L}_{\rm cons}$ & 2.0472 & 0.1217 & 1.2845 & 0.1346 \\
        \hline
    \end{tabularx}
    \label{tab:ablcons}
\end{table}

Besides, we explore the role of the consistency loss $\mathcal{L}_{\rm cons}$, which is mainly motivated to stabilize the training and generalize to real-world manga data.
As the result shown in Table~\ref{tab:ablcons}, it makes a significant improvement in the prediction stability but negligible accuracy gain on synthetic data. This is because the scale loss $\mathcal{L}_{\rm scl}$ can guarantee a sufficiently high accuracy already. In contrast, it indeed causes a stable numerical result on real-world data.

\subsection{Manga Restoration Quality}
\label{subsec:reconstruction}

\begin{figure}[!t]
    \centering
    \subfloat[Degraded manga]{\includegraphics[width=.25\linewidth]{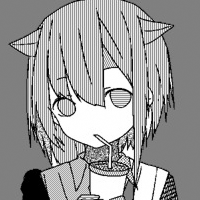}}
    \subfloat[Confidence map]{\includegraphics[width=.25\linewidth]{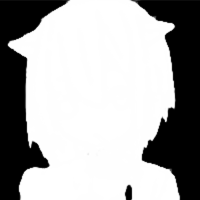}}
    \subfloat[Ours]{\includegraphics[width=.25\linewidth]{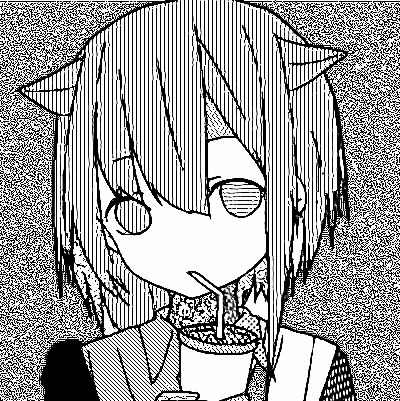}}
    \subfloat[Grouth truth]{\includegraphics[width=.25\linewidth]{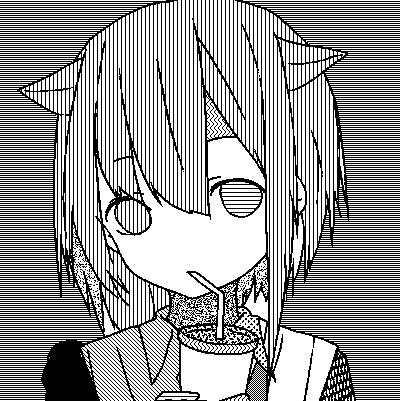}}
    \\
    \vspace{-0.5em}
    \caption{Manga restoration results with synthetic data. Some ambiguous regions are degraded into a plain regions which has no clue to restore the original version. }\vspace{-0.5em}
    \label{fig:certainty}
\end{figure}

\paragraph{Comparison with Baselines.}
After obtaining a proper scale factor, we can restore the manga image through the Manga Restoration Network (MR-Net). 
To evaluate the performance, we compare it with three typical super-resolution approaches: EDSR\cite{lim2017enhanced} which is a regression-based method, SRGAN\cite{ledig2017photo} which is a GAN-based method, and MetaSR\cite{hu2019meta} which is a method of arbitrary scale. The first two methods are trained with our dataset with a given scale factor ($\times2$). MetaSR and our MR-Net are trained with scale factors ranged in $[1,4]$. 
As most of screentones in our synthetic data lose their regularity when $\times3$ downscaled, we evaluate the performance on a synthetic dataset with the scale factors ranged in $[1,3]$. 
On those degraded images with scale factors $T\neq2$, the evaluation on EDSR\cite{lim2017enhanced} and SRGAN\cite{ledig2017photo} is performed by rescaling their results to the expected resolution. In particular, to avoid reference ambiguity, we quantitatively evaluate the restoration quality only on pattern-identifiable regions, as shown in Figure~\ref{fig:certainty}.

We report experiment results in Table~\ref{tab:compare} using PSNR, SSIM\cite{wang2004image} and SVAE. SVAE evaluates the screentone similarity between the generated results and the ground truth. It is achieved by comparing the ScreenVAE map \cite{xie2020manga} which is a continuous and interpolative representation for screentones.  
We can find that our model outperforms EDSR\cite{lim2017enhanced}, SRGAN\cite{ledig2017photo} and MetaSR\cite{hu2019meta} when the scale factor is various. 
Anyhow, our method can not achieve superiority over SRGAN\cite{ledig2017photo} at scale factor of 2 when SRGAN is trained to handle exactly the $\times 2$ scale while our model is trained for various target scales. However, as mentioned earlier, the model trained with fixed scale is infeasible to solve our problem in practical scenarios.
When combined with our SE-Net, MetaSR\cite{hu2019meta} can be roughly regarded as a reasonable baseline of our MR-Net. Note that our key concept is the scale-aware manga restoration, and the comparison with MetaSR that is provided with the ground-truth scale factors, is just to verify the structure effectiveness of the MR-Net.
The quantitative results shown in Table~\ref{tab:compare} illustrates the superiority of our MR-Net structure that adopts a flexible attention mechanism and discriminative restoration strategy.

\begin{figure*}[!t]
    \centering
    \includegraphics[width=.95\linewidth]{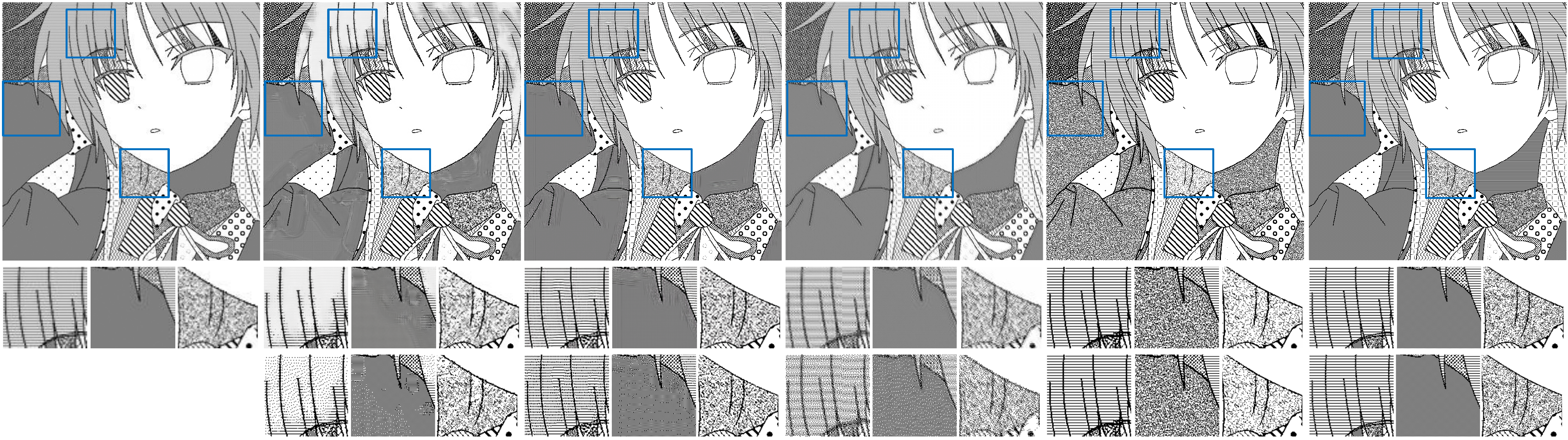}\\
    (a) Degraded manga \hspace{.05in} (b) EDSR\cite{lim2017enhanced} \hspace{.2in} (c) SRGAN\cite{ledig2017photo} \hspace{.2in} (d) MetaSR\cite{hu2019meta} \hspace{.4in} (e) Ours \hspace{.35in} (f) Grouth truth \hspace{.2in} \\ 
    \vspace{-.05in}
    \caption{Manga restoration results for synthetic data. Binarized results are shown on the bottom. EDSR\cite{lim2017enhanced}, SRGAN\cite{ledig2017photo} and MetaSR\cite{hu2019meta} may generate blurry screentones while our method can restore the bitonal nature. }
    \label{fig:synresults}
\end{figure*}

Figure~\ref{fig:synresults} compares the visual results on typical synthetic examples. 
Our method successfully restores the bitonal manga image from the degraded screentones.  
For the region where the information has totally lost after resampling, our result generates random but consistent bitonal screentones, leading to better visual results. 
Meanwhile, our screentones are consistent over regions and can be directly binarized with little information loss.

\begin{figure*}[!t]
    \centering
    \includegraphics[width=.95\linewidth]{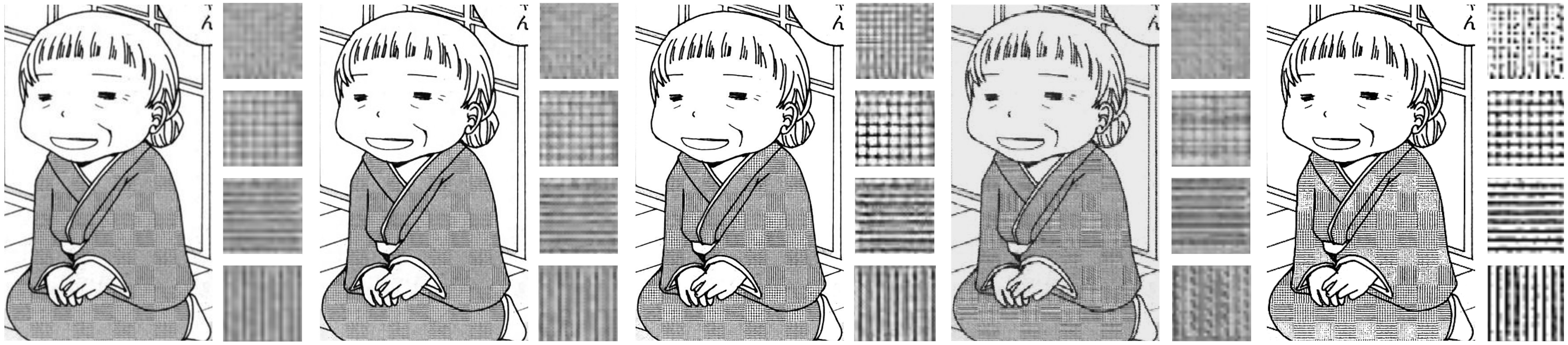}\\
    (a) Degraded manga \hspace{.05in} (b) EDSR\cite{lim2017enhanced}(200\%)  \hspace{.03in} (c) SRGAN\cite{ledig2017photo}(200\%) \hspace{.03in} (d) MetaSR\cite{hu2019meta}(127\%) \hspace{.2in} (e) Ours(127\%) \hspace{.2in} \\
    \vspace{-0.05in}
    \caption{Manga restoration results for real-world case. The screentones are shown on the right. The image comes from the Manga109 \cite{mtap_matsui_2017}. Akuhamu \textcopyright Arai Satoshi}
    \label{fig:realresults}
\end{figure*}

\begin{figure*}[!t]
    \centering
    \includegraphics[width=.95\linewidth]{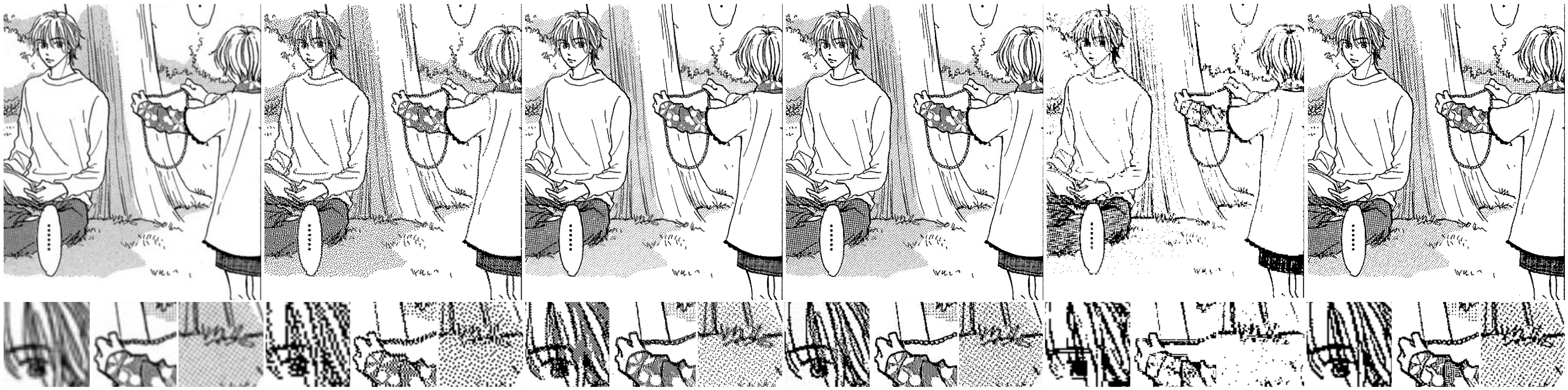}\\
    (a) Degraded manga \hspace{.05in} (b) Bitonal manga \hspace{.15in} (c) EDSR\cite{lim2017enhanced} \hspace{.15in} (d) SRGAN\cite{ledig2017photo} \hspace{.2in} (e) MetaSR\cite{hu2019meta} \hspace{.4in} (f) Ours \hspace{.4in} 
    \caption{Manga restoration results for real world case with bitonal nature. (b) is the binarized result under original resolution. (c) and (d) are restored under 200\% resolution while (e) and (f) are resotored under 150\% resolution. The image comes from the Manga109 \cite{mtap_matsui_2017}. HaruichibanNoFukukoro \textcopyright Yamada Uduki}
    \label{fig:realbresults}
\end{figure*}

\begin{table*}[!t]
    \centering
    \caption{Restoration accuracy of pattern-identifiable regions.}
    \vspace{-0.05in}
    \begin{tabular}{c|c|c|c|c|c|c|c|c|c}
        \hline
        Resolution & \multicolumn{3}{c}{$s_{\rm gt}=2$} & \multicolumn{3}{|c}{$s_{\rm gt}\in(1,2]$} & \multicolumn{3}{|c}{$s_{\rm gt}\in(2,3]$} \\
        \hline
        Metric & PSNR($\uparrow$) & SSIM($\uparrow$) & SVAE($\downarrow$) & PSNR & SSIM & SVAE & PSNR & SSIM & SVAE\\
        \hline
        EDSR\cite{lim2017enhanced} & 13.1695 & 0.6992 & 0.0318 & 13.9010 & 0.6206 & 0.0734 & 9.3550 & 0.2615 & 0.0717 \\
        \hline
        SRGAN\cite{ledig2017photo} & \textbf{14.8810} & \textbf{0.7829} & \textbf{0.0183} & 14.8987 & 0.8132 & 0.0353 & \textbf{12.5510} & 0.5418 & 0.0527 \\
        \hline
        MetaSR\cite{hu2019meta} & 10.029 & 0.2385 & 0.1006 & 12.3722 & 0.4032 & 0.0779 & 8.1153 & 0.1149 & 0.1011 \\
        \hline
        Ours & 11.5547 & 0.7101 & 0.0255 & \textbf{16.8054} & \textbf{0.8485} & \textbf{0.0222} & 12.0333 & \textbf{0.6214} & \textbf{0.0415} \\
        \hline
    \end{tabular}
    \label{tab:compare}
\end{table*}

\paragraph{Evaluation on Real-world Cases}
To validate the generalization, we test our method on some real-world cases (Manga109\cite{mtap_matsui_2017}). Results show that our method can restore visually pleasant results with clear screentones from the real-world degraded manga at the estimated upscaling resolution, as shown in Figure~\ref{fig:realresults}. As one may observe, our method can restore better results even with smaller resolutions. Since EDSR\cite{lim2017enhanced} and SRGAN\cite{ledig2017photo} are trained with specific scale factors, they may not restore the periodic information for some unseen screentones. MetaSR\cite{hu2019meta} failed to restore the bitonal nature. 
Our method is also friendly to do binarization, as shown in Figure~\ref{fig:realbresults}. We can see that although the regularity can be visually restored by EDSR\cite{lim2017enhanced} and SRGAN\cite{ledig2017photo} under a larger scale factor, the results cannot be directly binarized which may destroy the structures.
In contrast, our method can generate consistent screentones without destroying the structures. 

\begin{figure}[!t]
    \centering
    \subfloat[Degraded manga]{\includegraphics[width=.23\linewidth]{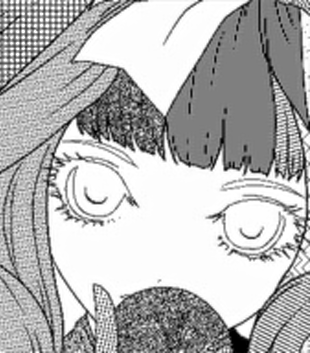}}\hspace{.01in}
    \subfloat[w/o $\mathcal{L}_{\rm pix}$]{\includegraphics[width=.23\linewidth]{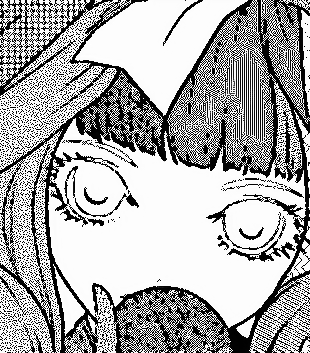}}\hspace{.01in}
    \subfloat[w/o $\mathcal{L}_{\rm conf}$]{\includegraphics[width=.23\linewidth]{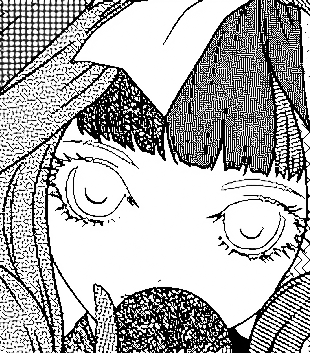}}\hspace{.01in}
    \subfloat[w/o $\mathcal{L}_{\rm bin}$]{\includegraphics[width=.23\linewidth]{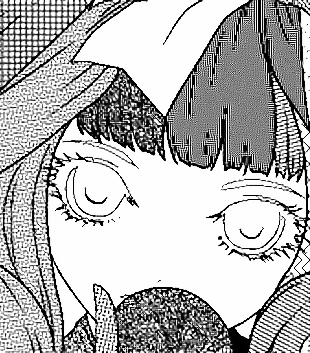}}\\ \vspace{-.1in}
    \subfloat[w/o $\mathcal{L}_{\rm itn}$]{\includegraphics[width=.23\linewidth]{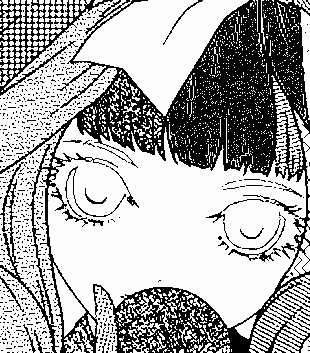}}\hspace{.01in}
    \subfloat[w/o $\mathcal{L}_{\rm hom}$]{\includegraphics[width=.23\linewidth]{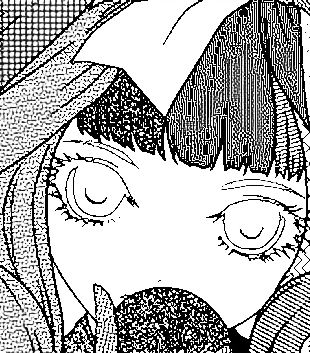}}\hspace{.01in}
    \subfloat[Ours]{\includegraphics[width=.23\linewidth]{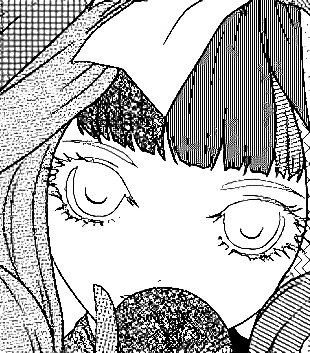}}\hspace{.01in}
    \subfloat[Target]{\includegraphics[width=.23\linewidth]{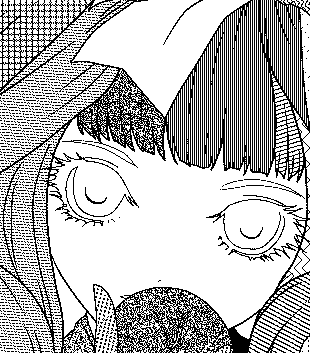}}
    \vspace{-.5em}
    \caption{Ablation study for individual loss term. }\vspace{-0.5em}
    \label{fig:ablation}
\end{figure}

\paragraph{Ablation Study for Individual Loss Terms.}
To verify the effectiveness of individual loss terms, we conduct ablation studies by visually comparing the generated output of different trained models without individual loss terms (Figure~\ref{fig:ablation}). The pixel loss $\mathcal{L}_{\rm pix}$ is the essential component to guarantee to restore the original image. Without the intensity loss $\mathcal{L}_{\rm itn}$, the pattern-agnostic regions may not follow the intensity constraint and thus generate undesired screentones. Meanwhile, the homogeneity loss $\mathcal{L}_{\rm itn}$ is important for generating consistent screentones in the pattern-agnostic regions.
In comparison, the combined loss can help the network to generate bitonal and consistent screentones for degraded manga images (Figure~\ref{fig:ablation} (g)). 

\begin{figure}[!t]
    \centering
    \subfloat[Degraded manga]{\includegraphics[width=.33\linewidth]{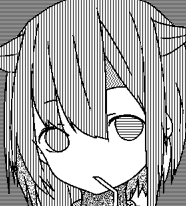}}
    \hspace{.02in}
    \begin{minipage}[b]{0.65\linewidth}
        \centering
        \subfloat[]{\includegraphics[width=.1\linewidth]{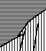}}\hspace{.02in}
        \subfloat[]{\includegraphics[width=.1\linewidth]{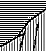}}\hspace{.02in}
        \subfloat[]{\includegraphics[width=.11\linewidth]{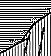}}\hspace{.02in}
        \subfloat[]{\includegraphics[width=.12\linewidth]{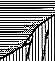}}\hspace{.02in}
        \subfloat[]{\includegraphics[width=.13\linewidth]{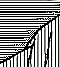}}\hspace{.02in}
        \subfloat[]{\includegraphics[width=.14\linewidth]{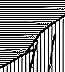}}\hspace{.02in}
        \subfloat[]{\includegraphics[width=.15\linewidth]{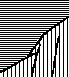}}\\
        \subfloat[]{\includegraphics[width=.16\linewidth]{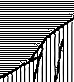}}\hspace{.02in}
        \subfloat[]{\includegraphics[width=.17\linewidth]{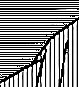}}\hspace{.02in}
        \subfloat[]{\includegraphics[width=.18\linewidth]{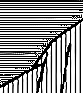}}\hspace{.02in}
        \subfloat[]{\includegraphics[width=.19\linewidth]{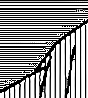}}\hspace{.02in}
        \subfloat[]{\includegraphics[width=.2\linewidth]{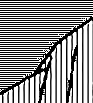}}
    \end{minipage}
    \vspace{-.5em}
    \caption{Manga restoration results under different resolutions. (b) is the degraded patch and (c)-(m) are the restored patches under resolutions ranges $100\%\sim200\%$. }
    \label{fig:range}
\end{figure}

\paragraph{Robustness to Restoration Scale.}
We argue that manga restoration requires to conduct at an appropriate resolution because of the target screentone is bitonal and usually regular. When the restorative resolution is not matched, the restored screentones may either cause blurry grayscale intensity or present irregular patterns. To verify the necessity of the scale estimation, we study the performance of the MR-Net with different resolutions.
As shown in Figure~\ref{fig:range}, only the results with ground-truth resolution (Figure~\ref{fig:range} (h)) achieve visually pleasant bitonal screentone patterns.

\subsection{Limitation and Discussion}
\label{subsec:limitation}

Our method still suffers from some limitations. 
Our model may fail to restore the bitonal screentones for some real-world cases. This is related to several aspects. 
Firstly, there are still gaps between synthetic data and real-world cases. Although our method improves the generalization in a semi-supervised manner, we may still fail to generalize to some unseen patterns. 
Secondly, in real-world applications, some degraded manga images are degraded by multiple times and have some other unconsidered operations, which are beyond the assumption of our problem setting. 

The pattern-agnostic regions are restored with screentones only under the condition of intensity constraint, which may cause perceptual inconsistency with the contextual screentones.
In our future works, we will try to generate controllable screentone types with user input. 
Xie et al.\cite{xie2020manga} proposed an effective point-wise representation of screentones, called ScreenVAE map. We may provide the ScreenVAE value as a hint for the pattern-agnostic regions and constrain the generated screentones to have similar ScreenVAE values, along with intensity constraint.

\section{Conclusion}
In this paper, we propose a deep learning method for manga restoration with learned scale factor. Our method first predicts a suitable scale factor for the low-resolution manga image. With the predicted scale factor, we further restore the high-resolution image which has bitonal and homogeneous screentones. Our method achieves high accuracy on synthetic data and can generate plausible results on real-world cases.

\section*{Acknowledgement}
This project is supported by Shenzhen Science and Technology Program (No.JCYJ20180507182410327) and The Science and Technology Plan Project of Guangzhou (No.201704020141).

{\small
\bibliographystyle{ieee_fullname}
\bibliography{egbib}

\begin{thebibliography}{10}\itemsep=-1pt

\bibitem{chang2004super}
Hong Chang, Dit-Yan Yeung, and Yimin Xiong.
\newblock Super-resolution through neighbor embedding.
\newblock In {\em Proceedings of the 2004 IEEE Computer Society Conference on
  Computer Vision and Pattern Recognition, 2004. CVPR 2004.}, volume~1, pages
  I--I. IEEE, 2004.

\bibitem{dong2015compression}
Chao Dong, Yubin Deng, Chen Change~Loy, and Xiaoou Tang.
\newblock Compression artifacts reduction by a deep convolutional network.
\newblock In {\em Proceedings of the IEEE International Conference on Computer
  Vision}, pages 576--584, 2015.

\bibitem{dong2015image}
Chao Dong, Chen~Change Loy, Kaiming He, and Xiaoou Tang.
\newblock Image super-resolution using deep convolutional networks.
\newblock {\em IEEE transactions on pattern analysis and machine intelligence},
  38(2):295--307, 2015.

\bibitem{freeman2002example}
William~T Freeman, Thouis~R Jones, and Egon~C Pasztor.
\newblock Example-based super-resolution.
\newblock {\em IEEE Computer graphics and Applications}, 22(2):56--65, 2002.

\bibitem{ge2018image}
Weifeng Ge, Bingchen Gong, and Yizhou Yu.
\newblock Image super-resolution via deterministic-stochastic synthesis and
  local statistical rectification.
\newblock In {\em SIGGRAPH Asia 2018 Technical Papers}, page 260. ACM, 2018.

\bibitem{goodfellow2014generative}
Ian Goodfellow, Jean Pouget-Abadie, Mehdi Mirza, Bing Xu, David Warde-Farley,
  Sherjil Ozair, Aaron Courville, and Yoshua Bengio.
\newblock Generative adversarial nets.
\newblock In {\em Advances in neural information processing systems}, pages
  2672--2680, 2014.

\bibitem{he2015delving}
Kaiming He, Xiangyu Zhang, Shaoqing Ren, and Jian Sun.
\newblock Delving deep into rectifiers: Surpassing human-level performance on
  imagenet classification.
\newblock In {\em Proceedings of the IEEE international conference on computer
  vision}, pages 1026--1034, 2015.

\bibitem{hu2019meta}
Xuecai Hu, Haoyuan Mu, Xiangyu Zhang, Zilei Wang, Tieniu Tan, and Jian Sun.
\newblock Meta-sr: A magnification-arbitrary network for super-resolution.
\newblock In {\em Proceedings of the IEEE Conference on Computer Vision and
  Pattern Recognition}, pages 1575--1584, 2019.

\bibitem{kim2016accurate}
Jiwon Kim, Jung Kwon~Lee, and Kyoung Mu~Lee.
\newblock Accurate image super-resolution using very deep convolutional
  networks.
\newblock In {\em Proceedings of the IEEE conference on computer vision and
  pattern recognition}, pages 1646--1654, 2016.

\bibitem{kingma2014adam}
Diederik~P Kingma and Jimmy Ba.
\newblock Adam: A method for stochastic optimization.
\newblock {\em arXiv preprint arXiv:1412.6980}, 2014.

\bibitem{lai2018fast}
Wei-Sheng Lai, Jia-Bin Huang, Narendra Ahuja, and Ming-Hsuan Yang.
\newblock Fast and accurate image super-resolution with deep laplacian pyramid
  networks.
\newblock {\em IEEE transactions on pattern analysis and machine intelligence},
  41(11):2599--2613, 2018.

\bibitem{ledig2017photo}
Christian Ledig, Lucas Theis, Ferenc Husz{\'a}r, Jose Caballero, Andrew
  Cunningham, Alejandro Acosta, Andrew Aitken, Alykhan Tejani, Johannes Totz,
  Zehan Wang, et~al.
\newblock Photo-realistic single image super-resolution using a generative
  adversarial network.
\newblock In {\em Proceedings of the IEEE conference on computer vision and
  pattern recognition}, pages 4681--4690, 2017.

\bibitem{li2017deep}
Chengze Li, Xueting Liu, and Tien-Tsin Wong.
\newblock Deep extraction of manga structural lines.
\newblock {\em ACM Transactions on Graphics (TOG)}, 36(4):1--12, 2017.

\bibitem{lim2017enhanced}
Bee Lim, Sanghyun Son, Heewon Kim, Seungjun Nah, and Kyoung Mu~Lee.
\newblock Enhanced deep residual networks for single image super-resolution.
\newblock In {\em Proceedings of the IEEE conference on computer vision and
  pattern recognition workshops}, pages 136--144, 2017.

\bibitem{liu2018non}
Ding Liu, Bihan Wen, Yuchen Fan, Chen~Change Loy, and Thomas~S Huang.
\newblock Non-local recurrent network for image restoration.
\newblock In {\em Advances in Neural Information Processing Systems}, pages
  1673--1682, 2018.

\bibitem{mtap_matsui_2017}
Yusuke Matsui, Kota Ito, Yuji Aramaki, Azuma Fujimoto, Toru Ogawa, Toshihiko
  Yamasaki, and Kiyoharu Aizawa.
\newblock Sketch-based manga retrieval using manga109 dataset.
\newblock {\em Multimedia Tools and Applications}, 76(20):21811--21838, 2017.

\bibitem{paszke2017automatic}
Adam Paszke, Sam Gross, Soumith Chintala, Gregory Chanan, Edward Yang, Zachary
  DeVito, Zeming Lin, Alban Desmaison, Luca Antiga, and Adam Lerer.
\newblock Automatic differentiation in pytorch.
\newblock 2017.

\bibitem{qu-2008-richness}
Yingge Qu, Wai-Man Pang, Tien-Tsin Wong, and Pheng-Ann Heng.
\newblock Richness-preserving manga screening.
\newblock {\em ACM Transactions on Graphics (SIGGRAPH Asia 2008 issue)},
  27(5):155:1--155:8, December 2008.

\bibitem{tao2018scale}
Xin Tao, Hongyun Gao, Xiaoyong Shen, Jue Wang, and Jiaya Jia.
\newblock Scale-recurrent network for deep image deblurring.
\newblock In {\em Proceedings of the IEEE Conference on Computer Vision and
  Pattern Recognition}, pages 8174--8182, 2018.

\bibitem{teed2020raft}
Zachary Teed and Jia Deng.
\newblock Raft: Recurrent all-pairs field transforms for optical flow.
\newblock In {\em European Conference on Computer Vision}, pages 402--419.
  Springer, 2020.

\bibitem{timofte2013anchored}
Radu Timofte, Vincent De~Smet, and Luc Van~Gool.
\newblock Anchored neighborhood regression for fast example-based
  super-resolution.
\newblock In {\em Proceedings of the IEEE international conference on computer
  vision}, pages 1920--1927, 2013.

\bibitem{tsubota2019synthesis}
Koki Tsubota, Daiki Ikami, and Kiyoharu Aizawa.
\newblock Synthesis of screentone patterns of manga characters.
\newblock In {\em 2019 IEEE International Symposium on Multimedia (ISM)}, pages
  212--2123. IEEE, 2019.

\bibitem{wang2017residual}
Fei Wang, Mengqing Jiang, Chen Qian, Shuo Yang, Cheng Li, Honggang Zhang,
  Xiaogang Wang, and Xiaoou Tang.
\newblock Residual attention network for image classification.
\newblock In {\em Proceedings of the IEEE conference on computer vision and
  pattern recognition}, pages 3156--3164, 2017.

\bibitem{wang2018esrgan}
Xintao Wang, Ke Yu, Shixiang Wu, Jinjin Gu, Yihao Liu, Chao Dong, Yu Qiao, and
  Chen Change~Loy.
\newblock Esrgan: Enhanced super-resolution generative adversarial networks.
\newblock In {\em Proceedings of the European Conference on Computer Vision
  (ECCV)}, pages 0--0, 2018.

\bibitem{wang2004image}
Zhou Wang, Alan~C Bovik, Hamid~R Sheikh, and Eero~P Simoncelli.
\newblock Image quality assessment: from error visibility to structural
  similarity.
\newblock {\em IEEE transactions on image processing}, 13(4):600--612, 2004.

\bibitem{wang2020deep}
Zhihao Wang, Jian Chen, and Steven~CH Hoi.
\newblock Deep learning for image super-resolution: A survey.
\newblock {\em IEEE Transactions on Pattern Analysis and Machine Intelligence},
  2020.

\bibitem{woo2018cbam}
Sanghyun Woo, Jongchan Park, Joon-Young Lee, and In So~Kweon.
\newblock Cbam: Convolutional block attention module.
\newblock In {\em Proceedings of the European conference on computer vision
  (ECCV)}, pages 3--19, 2018.

\bibitem{xie2020manga}
Minshan Xie, Chengze Li, Xueting Liu, and Tien-Tsin Wong.
\newblock Manga filling style conversion with screentone variational
  autoencoder.
\newblock {\em ACM Transactions on Graphics (SIGGRAPH Asia 2020 issue)},
  39(6):226:1--226:15, December 2020.

\bibitem{yang2008image}
Jianchao Yang, John Wright, Thomas Huang, and Yi Ma.
\newblock Image super-resolution as sparse representation of raw image patches.
\newblock In {\em 2008 IEEE conference on computer vision and pattern
  recognition}, pages 1--8. IEEE, 2008.

\bibitem{zhang2017learning}
Kai Zhang, Wangmeng Zuo, Shuhang Gu, and Lei Zhang.
\newblock Learning deep cnn denoiser prior for image restoration.
\newblock In {\em Proceedings of the IEEE conference on computer vision and
  pattern recognition}, pages 3929--3938, 2017.

\end{thebibliography}
}

\end{document}